\documentclass[aps,prc,nofootinbib]{revtex4}
\usepackage{eurosym}
\usepackage{amsmath,amssymb}
\usepackage{graphicx}
\usepackage{dcolumn}
\usepackage{bm}
\usepackage{color}
\usepackage{multirow}
\usepackage{float}
\usepackage{array}
\usepackage{dcolumn}
\usepackage{booktabs}
\usepackage[detect-all]{siunitx}
\usepackage{makecell}
\usepackage{xcolor}
\usepackage{yfonts}

\graphicspath{ {/Users/Nastia/Desktop/research_dis/papers_writing/xenes/nov_5} }

\newcolumntype{P}[1]{>{\centering\arraybackslash}p{#1}}
\begin{document}

\title{On a possible $^{3}_{\phi}$H hypernucleus with HAL QCD interaction
}
\author{I. Filikhin$^1$, R. Ya. Kezerashvili$^{2,3,4}$, and B. Vlahovic$^1$}
\affiliation{\mbox{$^{1}$North Carolina Central University, Durham, NC, USA} \\
$^{2}$New York City College of Technology, The City University of New York,
Brooklyn, NY, USA\\
$^{3}$The Graduate School and University Center, The City University of New
York, New York, NY, USA\\
$^{4}$Long Island University, Brooklyn, NY, USA}

\begin{abstract}
Within the framework of the Faddeev formalism in configuration space, we investigate bound states in the $\phi NN$ system with total isospin $T=0$ and $T=1$. The recently proposed lattice HAL QCD $\phi N$ potential in the 
 $^{4}S_{3/2}$ channel does not support either $\phi N$ or $\phi NN$ bound states.  The HAL QCD $\phi N$ potential in the $^{2}S_{1/2}$ channel suggests the bound states for $\phi N$ and $\phi NN (S=0)$ systems. However, the binding energies are highly sensitive to variations of the enhancement factor $\beta$, and the $\phi NN$ system is extremely strongly bound in the state $S=0$. Considering a spin-averaged potential 
for the state $S=1$ yields a bound state for $^3_\phi$H $(S=1)$ hypernucleus with the binding energy (BE) 14.9 MeV when $\beta = 6.9$. The evaluation of the BE for the $S=1$, $T=1$ three-body state results in 5.47 MeV.
Additionally, calculations using our approach confirm the bound states for the $\phi NN$ ($S=2,T=0$ and $S=1, T=1$) system previously predicted with the Yukawa-type potential motivated by the QCD van der Waals attractive force, mediated by multi-gluon exchanges.
\end{abstract}

\maketitle
\date{\today }

\section{Introduction}
Since the beginning of the new millennium, studying the composite system from two nucleons and $\Lambda$-, $\Xi$-, $\Omega$-hyperon or $\phi$-meson has
attracted intense research interest
in many theoretical works  \cite{Filikhin2000,FG2002,BSS,Bel2008,Sofi,GV15,GV2016,GVV16,FSV17,GV0,GV,Gibson2020,HiyamaXi2020,EF21,Zhang2022,GV2022,ESE2023,EA24}.
Unlike the case of
the $NN$ interactions, a $\phi$-meson nucleon interaction is not well determined due to an insufficient number of scattering data. It is one of the open and debated questions in the strangeness nuclear
physics concerns the possible existence of a $\phi N$ bound state.

The recent ALICE Collaboration measurement of the $\phi N$ correlation function \cite{ALICE2021} led
to the determination of the $\phi N$ channel scattering length with a large real part corresponding to an attractive interaction. This represents the first experimental evidence of the attractive strong interaction between a proton
and a $\phi$ meson. Interestingly, the absolute value of the obtained scattering length 
is much larger than what has been measured in earlier $\phi$-meson photoproduction experiments \cite{Chang2008,Strakovsky2020}.

It has been suggested by Brodsky, Schmidt, and de Teramond \cite{Brodsky1990} that the QCD van der Waals
interaction, mediated by multi-gluon exchanges, is dominant
when the interacting two color singlet hadrons have no common
quarks. 
Assuming that the attractive QCD van der Waals force dominates the $\phi N$ interaction since the
$\phi$-meson is almost a pure $s\bar s$ state, following \cite{Brodsky1990}, Gao, et al. \cite{G2001} suggested a Yukawa-type attractive potential. Using the variational method, they predicted  a binding energy (BE) of 1.8 MeV for the $\phi$-$N$ system.
In \cite{ALICE2021} the data are employed to constrain the parameters
of phenomenological Yukawa-type potentials. The resulting values for the Yukawa-type potential, $V_{\phi N}(r)=-A e^{-\alpha r}/r$, yields $A = 0.021\pm \text{0.009(stat.)}\pm \text{0.006(syst.)}$ and $\alpha = 65.9\pm  \text{38.0(stat.)}\pm \text{17.5(syst.)}$ MeV. Predictions of possible $\phi N$ bound states employing the same kind of potential with
parameters $A = 1.25$ and $\alpha = 600$ MeV \cite{G2001} are therefore incompatible with measurement \cite{ALICE2021}.

Recently, Lyu et al. \cite{Lyu22} presented the first results on the interaction between the $\phi$-meson and the nucleon based on the
(2+1)-flavor lattice QCD simulations with nearly physical quark masses.
The HAL QCD potential is obtained from first principles
(2 +1)-flavor lattice QCD simulations in a large spacetime
volume, $L^4 = $(8.1  \text{fm})$^4$, with the isospin-averaged masses of $\pi$,
$K$, $\phi$, and $N$ as 146, 525, 1048, and 954 MeV, respectively,
at a lattice spacing,
$a = 0.0846$ fm. Let us mention that such simulations together with the HAL
QCD method 
enable one to extract the $YN$ and $YY$
interactions with multiple strangeness, e.g., $\Lambda\Lambda$, $\Xi N$ \cite{Sasaki2020},
$\Omega N$  \cite{Iritani2019}, $\Omega\Omega$ \cite{OmegaOmega2018}, and $\Xi N$ \cite{HiyamaXi2020}. 
Using the HAL QCD method, based on
the spacetime correlation of the $\phi N$ system in the spin 3/2 channel, the authors suggested fits of the lattice QCD potential in the $^{4}S_{3/2}$ channel.
In the following, we employ the spectroscopic notation
$^{2s+1}S_{J}$ to classify the $S$-wave $\phi N$ interaction, where
$s$ and $J$ stand for total spin, and total angular momentum.
It was found that the simple fitting functions such as the
Yukawa form cannot reproduce the lattice data \cite{Lyu22}. The lattice calculations for the
$\phi N$ interaction in the $^{4}S_{3/2}$ channel are used in \cite{Chizzali2024} to constrain the spin 1/2 counterpart ($^{2}S_{1/2}$) from the fit of the experimental $\phi N$ correlation function measured by the ALICE Collaboration \cite{ALICE2021}.

The mesic $\phi NN$ system is considered in the framework of Faddeev equations in the differential form \cite{BSS}, using the variational folding method \cite{Bel2008}, and a two-variable integro-differential equation describing bound
systems of unequal mass particles \cite{Sofi}. Calculations were employed $\phi N$ potential from \cite{G2001}. 
The binding energy of $\phi d$ hypernucleus was calculated by employing HAL QCD potential \cite{Lyu22} using the Schrodinger equation for Faddeev components 
expanded in terms of hyperspherical functions \cite{EA24}. The binding energies reported in Refs. \cite{BSS,Bel2008,EA24} are in the range of $\sim 6-39$ MeV.

Motivated by the above discussion and the availability of newly suggested HAL QCD potentials in the $^{2}S_{1/2}$ and $^{4}S_{3/2}$ channels with a minimal and maximal spin, respectively, we present calculations for the binding energy for the $\phi N$ and $\phi NN$ in the framework of the Faddeev equations in configuration space. We compare our results with other calculations as well.

The $\phi NN$ represent a three-particle system.
The three-body problem can be solved in the framework of the Schr\"{o}dinger
equation or using the Faddeev approach in the momentum \cite{Fad,Fad1} or
configuration \cite{Noyes1968,Noyes1969,Gignoux1974,FM,K86} spaces. With
regards to the Faddeev
equations in the configuration space, Jacobi coordinates are
introduced to describe the $\phi NN$ system. The mass-scaled Jacobi coordinates
$\mathbf{x}_{i}$ and $\mathbf{y}_{i}$ are expressed via the particle
coordinates $\mathbf{r}_{i}$ and masses $m_{i}$ in the following form:
\begin{equation}
\mathbf{x}_{i}=\sqrt{\frac{2m_{k}m_{l}}{m_{k}+m_{l}}}(\mathbf{r}_{k}-\mathbf{%
r}_{l}),\qquad \mathbf{y}_{i}=\sqrt{\frac{2m_{i}(m_{k}+m_{l})}{%
m_{i}+m_{k}+m_{l}}}(\mathbf{r}_{i}-\frac{m_{k}\mathbf{r}_{k}+m_{l}\mathbf{r}%
_{l})}{m_{k}+m_{l}}).  \label{Jc}
\end{equation}%
The orthogonal transformation between three different sets of
the Jacobi coordinates has the form:
\begin{equation}
\left(
\begin{array}{c}
\label{tran}\mathbf{x}_{i} \\
\mathbf{y}_{i}%
\end{array}%
\right) =\left(
\begin{array}{cc}
C_{ik} & S_{ik} \\
-S_{ik} & C_{ik}%
\end{array}%
\right) \left(
\begin{array}{c}
\mathbf{x}_{k} \\
\mathbf{y}_{k}%
\end{array}%
\right) ,\ \ C_{ik}^{2}+S_{ik}^{2}=1, \quad k\neq i, \quad C_{ii}=1,
\end{equation}%
where
\begin{equation*}
C_{ik}=-\sqrt{\frac{m_{i}m_{k}}{(M-m_{i})(M-m_{k})}},\quad S_{ik}=(-1)^{k-i}%
\mathrm{sign}(k-i)\sqrt{1-C_{ik}^{2}}.
\end{equation*}%
Here, $M$ is the total mass of the system. Let us definite the
transformation $h_{ik}(\mathbf{x},\mathbf{y})$ based on Eq. (\ref{tran}) as
\begin{equation}
h_{ik}(\mathbf{x},\mathbf{y})=\left(C_{ik} \mathbf{x}+ S_{ik}\mathbf{y},
-S_{ik}\mathbf{x}+ C_{ik}\mathbf{y} \right).  \label{Trans}
\end{equation}
 In the Faddeev
method in configuration space, alternatively to the finding the wave
function of the three-body system using the Schr\"{o}dinger equation, the
total wave function is decomposed into three components \cite%
{Noyes1968,FM,K86}:
$\Psi (\mathbf{x}_{1},\mathbf{y}_{1})=\Phi_{1}(\mathbf{x}_{1},\mathbf{y}%
_{1})+\Phi_{2}(\mathbf{x}_{2},\mathbf{y}_{2})+\Phi_{3}(\mathbf{x}_{3},%
\mathbf{y}_{3}). $
Each component depends on the corresponding coordinate set,
which are expressed in terms of the chosen set of mass-scaled Jacobi
coordinates. The transformation (\ref{Trans}) allows us to write the Faddeev equations as a system of differential equations for each  $\Phi_i(\mathbf{x}_{i},\mathbf{y}_{i})$ component in
compact form. The components $\Phi_i(\mathbf{x}_{i},\mathbf{y}_{i})$ satisfy
the Faddeev equations \cite{FM} that can be written in the coordinate
representation as:
\begin{equation}
(H_{0}+V_{i}(C_{ik}\mathbf{x})-E)\Phi_i(\mathbf{x},\mathbf{y})=-V_{i}(C_{ik}%
\mathbf{x})\sum_{l\neq i}\Phi_l(h_{il}(\mathbf{x},\mathbf{y})).  \label{e:1}
\end{equation}
Here $H_{0}=-(\Delta _{\mathbf{x}}+\Delta _{\mathbf{y}})$ is the
kinetic energy operator with $\hbar ^{2}=1$ and $V_{i}(\mathbf{x})$ is the
interaction potential between the pair of particles $(kl)$, where $k,l\neq i$.

The system of Eqs. (\ref{e:1}) written for three nonidentical particles can be
reduced to a simpler form for a case of two identical particles. 
The Faddeev
equations in configuration space for a three-particle system with two identical
particles are given in our previous studies \cite{Kez2018PL,FilKez2018,KezPRD2020}%
. In the case of the $\phi NN$ system, the total wave function of the
system is decomposed into the sum of the Faddeev components $\Phi_1$ and $%
\Phi_2$ corresponding to the $(NN)\phi$ and $(\phi N)N$ types of rearrangements:
$\Psi =\Phi_1+\Phi_2-P\Phi_2$,
where $P$ is the permutation operator for two identical particles.
Therefore, the set of the Faddeev equations (\ref{e:1}) is rewritten as
follows \cite{K86}:
\begin{equation}
\begin{array}{l}
{(H_{0}+V_{NN}-E)\Phi_1=-V_{NN}(\Phi_2-P\Phi_2)}, \\
{(H_{0}+V_{\phi N}-E)\Phi_2=-V_{\phi N}(\Phi_1-P\Phi_2)}.%
\end{array}
\label{GrindEQ__1_}
\end{equation}
In Eqs. (\ref{GrindEQ__1_}) ${V_{NN}}$ and ${V_{\phi N}}$ are the interaction
potentials between two nucleons and the $\phi$-meson and nucleon, respectively. The
spin-isospin variables of the system can be represented by the
corresponding basis elements. After the separation of the variables, one can
define  the coordinate part the $\Psi^{R}$ of the wave function $\Psi =\xi
_{isospin}\otimes \eta _{isospin}\otimes \Psi ^{R}$. The
details of our method for the solution of the system of differential equations (\ref{GrindEQ__1_}) are given in \cite{KezPRD2020,KezJPG2024,KezJPG2016}.

In Ref. \cite{Lyu22}, the interaction between the $\phi $-meson and the
nucleon is studied based on the ($2+1$)-flavor lattice QCD simulations with
nearly physical quark masses. The authors found that the $\phi N$ correlation
function is mostly dominated by the elastic scattering states in the $%
^{4}S_{3/2}$ channel without significant effects from the two-body $\Lambda
K(^{2}D_{3/2})$ and $\Sigma K(^{2}D_{3/2})$ and the three-body open channels
including $\phi N\rightarrow {\Sigma ^{\ast }K,\Lambda (1405)K}\rightarrow {%
\Lambda \pi K,\Sigma \pi K}$. The fit of the lattice QCD potential by the
sum of two Gaussian functions for an attractive
short-range part and a two-pion exchange tail at long distances with an
overall strength proportional to $m_{\pi }^{4n}$ \cite{Kreinm4}, has the
following functional form in the $^{4}S_{3/2}$ channel with the maximum spin $3/2$ \cite{Lyu22}:
\begin{equation}
V_{\phi N}^{3/2}(r)=\sum_{j=1}^{2}a_{j}\exp \left[ -\left( \frac{r}{b_{j}}%
\right) ^{2}\right] +a_{3}m_{\pi }^{4}F(r,b_{3})\left( \frac{e^{-m_{\pi }r}}{%
r}\right) ^{2},  \label{HALQCD}
\end{equation}%
with the Argonne-type form factor \cite{Wiringa95}
\begin{equation}
F(r,b_{3})=(1-e^{-r^{2}/b_{3}^{2}})^{2}.  \label{Ffactor}
\end{equation}%
For comparison the lattice QCD $\phi N$ potential is also parameterized
using three Gaussian functions \cite{Lyu22}:
\begin{equation}
V_{G\phi N}(r)=\sum_{j=1}^{3}a_{j}\exp \left[ -\left( \frac{r}{b_{j}}\right)
^{2}\right] .
\end{equation}%
The HAL QCD potential in the $^{2}S_{1/2}$ channel with a minimum spin of $1/2$ \cite{Chizzali2024} has a much stronger attractive
$\beta$-enhanced short-range part 
and the same two-pion exchange
long-range tail as in the $^{4}S_{3/2}$ channel. The real part of the potential
in the $^{2}S_{1/2}$ channel reads \cite{Chizzali2024}
\begin{equation}
V_{\phi N}^{1/2}(r)=\beta \left(
a_{1}e^{-r^{2}/b_{1}^{2}}+a_{2}e^{-r^{2}/b_{2}^{2}}\right) +a_{3}m_{\pi
}^{4}F(r,b_{3})\left( \frac{e^{-m_{\pi }r}}{r}\right) ^{2},
\label{9}
\end{equation}%
where the factor $\beta =$6.9$_{-0.5}^{+0.9}$(stat.)$_{-0.1}^{+0.2}$(syst.).
The other values of the parameters are common in both $^{4}S_{3/2}$ and $%
^{2}S_{1/2}$ channels \cite{Chizzali2024}. The imaginary part of $\phi N$ potential related to
the 2nd-order kaon exchange and corresponds to absorption processes. A
proportionality coefficient for this part is $\gamma =$0.0$_{-3.6}^{+0.0}$%
(stat.)$_{-0.18}^{+0.0}$(syst.)
\cite{Chizzali2024}.
\begin{table}[!ht]
\begin{center}
\caption{The parameters for the $\phi N$ potential in the $^{4}S_{3/2}$ channel with statistical errors are quoted in parentheses. For the $a_{3}m_{\pi }^{4n}$ column, $n=1$ and $n=0$ for $V_{\phi N}^{3/2}$ and $V_{G\phi N}^{3/2}$, respectively \cite{Lyu22}. 
The parameters for the singlet and triplet $NN$ interaction for MT potential \cite{Malfliet1969,MTcorr}. }
\label{tpp}
\begin{tabular}{ccccccc}
\hline  \noalign{\smallskip}
& \multicolumn{5}{c}{$_{\phi N}$ potential in the $^{4}S_{3/2}$ channel \cite{Lyu22} and in the $^{2}S_{1/2}$ channel \cite{Chizzali2024} } \\ \hline   \noalign{\smallskip}
& $a_{1},$ MeV & $a_{2},$ MeV & $a_{3}m_{\pi }^{4n},$ MeV fm$^{2n}$ & $b_{1},
$ fm & $b_{2},$ fm & $b_{3},$ fm \\ \cline{1-7}  \noalign{\smallskip}
$V_{\phi N}^{3/2}$ & -371(27) & -119(39)  & -97(14)  & 0.13(1) & 0.30(5) & 0.63(4)
\\
$V_{G\phi N}$ & -371(19) & -50(35) & -31(53)  & 0.15(3) & 0.66(61) & 1.09(41)
\\ \hline  \noalign{\smallskip}
& \multicolumn{5}{c}{Singlet $^{1}S_{0}$ and triplet $^{3}S_{1}$ $NN$
potential \cite{Malfliet1969,MTcorr}}  \\ \cline{2-6}  \noalign{\smallskip}
& $I,J$ & $V_{r},$ MeV & $V_{a},$ MeV & $\mu _{1},$ fm$^{-1}$  & $\mu _{2},$
fm$^{-1}$ &  \\ \cline{2-6}  \noalign{\smallskip}
& 1,0 & -521.959 & 1438.72 & 1.55 & 3.11 &  \\
& 0,1 & -626.885 & 1438.72 & 1.55 & 3.11 &   \\ \cline{2-6}
\end{tabular}
\end{center}
\end{table}


We present the results of calculations for the feasibility of expected bound states for $\phi N$ and $\phi NN$ systems. For calculations of the BEs of these systems, we use the HAL QCD $\phi N$  potential in the $^{4}S_{3/2}$ and  $^{2}S_{1/2}$ channels  with the maximum and minimum spins, respectively. 
We employ the same $NN$ MT-I-III potential \cite{Malfliet1969,MTcorr} as in \cite{BSS,Bel2008,Sofi,EA24} for the comparison of the results. The input parameters for potentials are listed in Table \ref{tpp}. For comparison, we also perform BE calculations  for $\phi N$ and $\phi NN$ systems with previously suggested Yukawa-type $\phi N$ potential with parameters from \cite{G2001} and \cite{ALICE2021}.

\begin{figure}[t]
\begin{center}
\includegraphics[width=19pc]{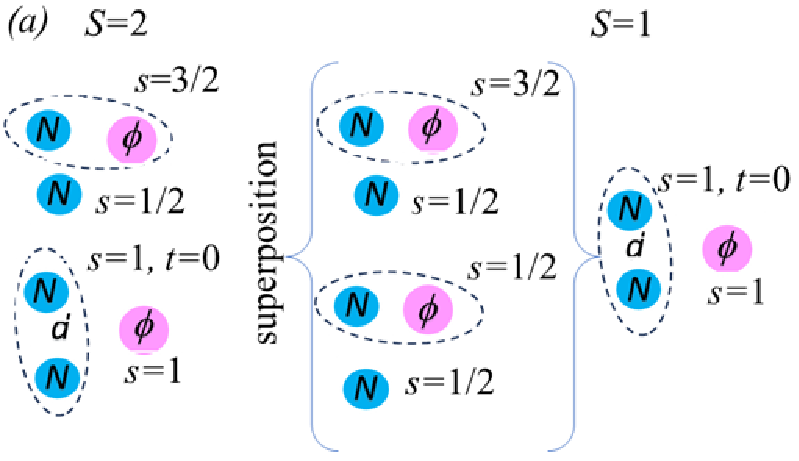}
\includegraphics[width=19pc]{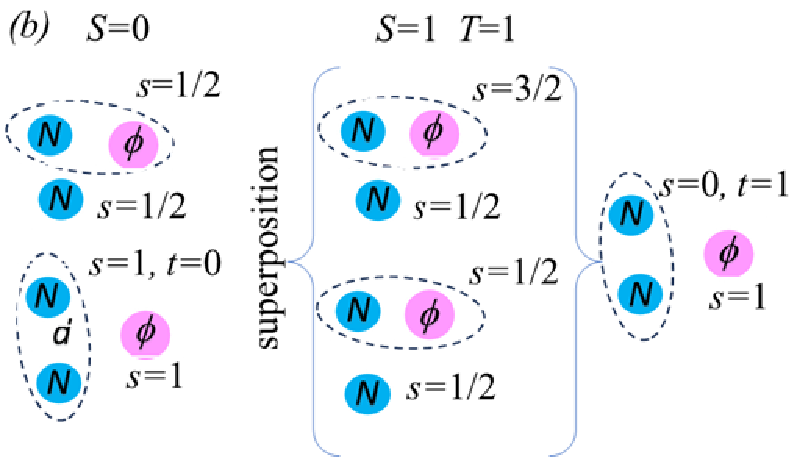}
\end{center}
\caption{Spin-isospin configurations in the $\phi NN$ system: (a) $S=2$, $T=0$ and $S=1$, $T=0$, (b) $S=0$, $T=0$ and $S=1$, $T=1$.  The channels $(\phi N)N$ and $\phi (NN)$ are shown
} \label{fig:05}
\end{figure}

The spin configurations of the $\phi NN$ system are illustrated in Fig. \ref{fig:05}$(a)$.
Here, we present two configurations for isospin state $T=0$ which means that the considered system includes the deuteron, $d$, which is corresponding to the $NN(s=1)$ state.
There are two different components of the $\phi N$ potential.
For calculations for the $S=1$ state we used an averaged over spin variables potential.
To acquire the overall $\phi N$ potential, the spin-averaged interaction for the $^{4}S_{3/2}$ and  $^{2}S_{1/2}$ channel potentials is defined as \cite{Chizzali2024}
\begin{equation}
\label{Mix}
\bar V_{\phi N}=\frac{1}{3}V_{\phi N}^{1/2} + \frac{2}{3}V_{\phi N}^{3/2}.
\end{equation}
According to Eq. (\ref{Mix}), the configuration $S=1$ becomes to the configuration $S=2$ when components of the  $\phi N$
potential are equal. For example, it can be  the $3/2$ $\phi N$ component.
The configuration for $S=0$ and $S=1$, $T=1$ states are presented in Fig. \ref{fig:05}$(b)$.

First, let us consider the $\phi N$ system. Results of calculations for the two-body binding energy, $B_{2}$, scattering length, $a_{\phi N}$, and effective radius, $r_{\phi N}$, for $\phi N$ are presented in Table \ref{Rt33} for the $^{4}S_{3/2}$ and $^{2}S_{1/2}$ channels.  Although the HAL QCD $\phi N$ potential in the $^{4}S_{3/2}$ channel is found to be attractive for all distances and reproduces a two-pion exchange tail at long distances, no bound $\phi N$ state is found with this interaction. The $\phi N$ system is strongly bound with the HAL QCD potential in the $^{2}S_{1/2}$ channel with the reasonable scattering length when the short-range attractive part is enchanted with factor $\beta = 6.9$ suggested in \cite{Chizzali2024}. Let us mention that the $^{2}S_{1/2}$ state binding energy is very sensitive to the variation of $\beta$ within the statistical and systematic error margins reported in \cite{Chizzali2024}.

\begin{table}[h!]
\caption{The scattering length $a_{\phi N}^{3/2}$ and $a_{\phi N}^{1/2}$, effective radius $r_{\phi
N}^{3/2}$ and $r_{\phi N}^{1/2}$ in fm and binding energies  $B_{2}^{3/2}$ and $B_{2}^{1/2}$ in MeV for $\phi N$ in the $s=3/2$ and $s=1/2$ spin states, respectively, $B_{3}$ in MeV is the binding energy of the $\phi d$ or $\phi NN$.  $\beta$ is the scaling factor for the attractive short-range part of $V_{\phi N}^{1/2}$ potential (see Eq. (\ref{9})). The "UNB" indicates that no bound state is found. The bound energy $B^{\phi NN}_{3}$ of the $\phi NN$ system ($S=1$, T=1) is shown in  parentheses. }
\label{Rt33}
\begin{tabular}{ccccccccccc}
\hline
 \noalign{\smallskip}
$\phi N$ potential &  $\beta$ & $a_{\phi N}^{3/2}$ & $a_{\phi N}^{1/2}$ & $r_{\phi
N}^{3/2}$ & $r_{\phi N}^{1/2}$ & $B_{2}^{3/2}$ & $B_{2}^{1/2}$ & $B^{\phi NN}_{3}(S=2)$ & $B^{\phi NN}_{3}(S=1)$ & $B^{\phi NN}_{3}(S=0)$\\
 \noalign{\smallskip}\hline \noalign{\smallskip}
-$A\frac{e^{-\alpha r}}{r}$ \cite{ALICE2021} &--  & $-1.13$ & -- & 36.4 &--  & UNB &--  & UNB& -- & --\\
-$A\frac{e^{-\alpha r}}{r}$ \cite{G2001} & -- & 2.38 & -- & 0.17 &--  & 9.40 & -- & 38.04  &-- (22.42)& --
\\
$V_{\phi N}^{3/2}$ $(^{4}S_{3/2})$ \cite{Lyu22} & -- & $-1.37$ &--  & 2.42 & -- & UNB &--  & UNB&--&-- \\
$V_{G\phi N}^{3/2}$ $(^{4}S_{3/2})$ \cite{Lyu22} & -- & $-1.36$ &--  & 2.04 &--  & UNB &--  & UNB&--  &-- \\
$(\frac{1}{3}V_{\phi N}^{1/2}+\frac{2}{3}V_{\phi N}^{3/2})$\cite{Chizzali2024}&
6.9\cite{Chizzali2024}& -1.37 & 1.5 & 2.24 &  $\sim 0$ & UNB & 27.7 &--&14.90 (5.47)& --\\
$V_{\phi N}^{1/2}$\cite{Chizzali2024}&
6.9\cite{Chizzali2024}&-- & 1.5 & -- &  $\sim 0$ & -- & 27.7 & --&--& 64.13\\
\noalign{\smallskip}\hline   \noalign{\smallskip}
$(\frac{1}{3}V_{\phi N}^{1/2}+\frac{2}{3}V_{\phi N}^{3/2})$& $5.0$ & -1.37 & 8 &2.24  & $0.7$ & UNB  & $0.7$
 &--&$11.37$& -- \\
$V_{\phi N}^{1/2}$& $5.0$ & -- & 8 &--  & $0.7$ & -- & $0.7$
 &--&--& 18.56 \\
$(\frac{1}{3}V_{\phi N}^{1/2}+\frac{2}{3}V_{\phi N}^{3/2})$ & $6.0$ & -1.37  & 2.5 & 2.24 & $0.3$ & UNB & $%
8.81$ &-- & $13.09$& --\\
$V_{\phi N}^{1/2}$ & $6.0$ & -- & 2.5 &-- & $0.3$ & --&
8.81 &-- & -- & 37.11\\
 $(\frac{1}{3}V_{\phi N}^{1/2}+\frac{2}{3}V_{\phi N}^{3/2})$&
6.9& -1.37 & 1.5 & 2.24 &  $\sim 0$ & UNB & 27.7 &--&14.90 & --\\
$V_{\phi N}^{1/2}$&
6.9&-- & 1.5 & -- &  $\sim 0$ & -- & 27.7 & --&--& 64.13\\
$(\frac{1}{3}V_{\phi N}^{1/2}+\frac{2}{3}V_{\phi N}^{3/2})$& 8.0 &  -1.37 & 1 & 2.24 & $\sim 0$ & UNB &
69.85 &--& $17.52$ & --\\
$V_{\phi N}^{1/2}$& 8.0 & -- & 1 & -- & $\sim 0$ &--  &
69.85 &--& -- & 113.7\\
\hline
\end{tabular}
\end{table}
In Table \ref{Rt33} we present the numerical results for the $\phi NN$ system obtained with the HAL QCD interactions and a Yukawa-type potential with parameterizations from \cite{G2001} and \cite{ALICE2021}. The calculations of the BEs  with the Yukawa-type potential  motivated by the QCD van der Waals attractive force mediated by multi-gluon exchanges, led to the same results as previously reported in \cite{BSS,Bel2008,Sofi}.
Our calculations indicate that neither HAL QCD interaction in the $^{4}S_{4/2}$ channel nor the Yukawa type interaction with parameters \cite{ALICE2021} do not support the existence of the $S=2$ bound state. Thus, the HAL QCD interaction in the $^{4}S_{3/2}$ channel with the maximum spin $3/2$ suggests
no bound state for $^3_\phi$H hypernucleus, in contrast to the binding energy range reported in \cite{EA24}, which is $6.7-7.3$ MeV. Results obtained for the BE of $\phi NN$ 22.42 MeV and 38.04 MeV ($t=0$) in the framework of our approach utilizing the Yukawa-type $\phi N$ potential \cite{G2001} and the singlet and triplet spin $NN$ interaction \cite{Malfliet1969}, respectively, confirm calculations \cite{BSS,Sofi} and are in good agreement within $\pm 1.5$ MeV.

Based on our calculations, the HAL QCD interaction in the $^{4}S_{3/2}$ channel does not provide enough attractiveness to bind a $\phi$-meson onto a nucleon or deuteron to form a bound state. Conversely, employing the HAL QCD $\phi N$ interaction in the $^{2}S_{1/2}$ channel with minimal spin $1/2$ results in the bound $\phi NN$, although the BE is highly sensitive to the variation of the factor $\beta$, and the $\phi NN$ system is extremely strongly bound in the state $S=0$. Employing the spin-averaged potential (\ref{Mix}), we consider both the HAL QCD potentials in the $^{2}S_{1/2}$ and $^{4}S_{3/2}$ channels when the factor $\beta = 6.9$. This leads to the numerical value of the binding energy 14.9 MeV for the $^3_\phi$H hypernucleus in the spin state $S=1$.
Changing the $\beta$ factor to  $\beta = 6.0$, we obtained for the  $\phi NN$ BE 13.09 MeV, albeit with a larger scattering length.
It is important to note that varying the $\beta$ factor within the margin of the error leads to larger and less realistic BEs, especially for the $S=0$ state
as shown in Table \ref{Rt33}.

In conclusion, we employ  the HAL QCD $\phi N$ potential in the $^{2}S_{1/2}$ and $^{4}S_{3/2}$ channels with the maximum and minimum spin, respectively, in the framework of Faddeev equations in configuration space to evaluate the binding energy of the $\phi NN$ system. The HAL QCD  $\phi N$ potential in the $^{4}S_{3/2}$ channel does not support bound states for either $\phi N$ or $\phi NN$, although it exhibits attraction.
Conversely, employing the HAL QCD $\phi N$ potential in the $^{2}S_{1/2}$ channel yields bound states for both $\phi N$ and $\phi NN$. The binding energies of
these systems are notably sensitive to variations in the enhancement of the short-range attractive part, parameterized
by the factor $\beta$. Considering both potentials, we find binding energies of 5.47 MeV and 14.9 MeV for the states  $S=1$, $T=0$ and $S=1$, $T=1$ (with singlet and triplet components of the $NN$ MT I-III potential), respectively, when $\beta = 6.9$. Our calculations confirm the existence of $S=2$ bound states for the $\phi NN$ system previously predicted within the Faddeev equations in the differential form \cite{BSS} and theoretical formalism \cite{Sofi} where utilized $\phi N$ potential \cite{G2001}. The presented analysis demonstrates the possible existence of $^{3}_{\phi}$H hypernucleus.

\section*{Acknowledgments}

This work is supported by the City University of New York PSC CUNY Research Award \# 66109-00 54 and
 US National Science Foundation HRD-1345219 award, the DHS (summer research team), and the Department of Energy/National Nuclear Security Administration Award Number DE-NA0004112.

\end{document}